# Coaxial Atomic Force Microscope Tweezers

K. A. Brown, J. A. Aguilar,[a] and R. M. Westervelt[b]

Harvard University School of Engineering and Applied Science and Department of Physics, Cambridge, Massachusetts 02138

**Abstract**

We demonstrate coaxial atomic force microscope (AFM) tweezers that can trap and place small objects using dielectrophoresis (DEP). An attractive force is generated at the tip of a coaxial AFM probe by applying a radio frequency voltage between the center conductor and a grounded shield; the origin of the force is found to be DEP by measuring the pull-off force *vs.* applied voltage. We show that the coaxial AFM tweezers (CAT) can perform three dimensional assembly by picking up a specified silica microsphere, imaging with the microsphere at the end of the tip, and placing it at a target destination.

---

[a] Currently at AGH University of Science and Technology

[b] Electronic Mail: westervelt@seas.harvard.edu

The assembly of small objects is important for bottom-up construction of nanostructures as well as the manipulation of cells and biological systems. Manipulation of objects with an atomic force microscope (AFM) is a topic of current interest, because an AFM probe can be positioned with nm precision, image topography, and locally measure pN-scale forces.[1] In addition to lithographic techniques[2] such as Dip-Pen Nanolithography,[3] AFMs have been used to manipulate objects by sliding them on surfaces[4,5] and by using two coordinated AFMs probes.[6] Coaxial AFM probes that concentrate an electric field near their tip have been used for scanning near-field microwave[7] and optical microscopy.[8] Another approach to assembly is using dielectrophoresis (DEP) which is widely used to manipulate cells[9] and nanoscale objects.[10] A triaxial AFM tip was recently proposed to for non-contact manipulation of nanoparticles using DEP.[11]

In this letter, we propose and demonstrate coaxial AFM tweezers (CAT), a tool for 3D manipulation of dielectric objects. A CAT tip consists of two coaxial electrodes: a conducting AFM tip separated from a ground shield by an insulating layer, as shown in Fig. 1(a). By applying a radio frequency (RF) voltage between the conducting tip and the ground shield, we create an electric field that is strongly peaked at the end of the tip, as shown in Fig. 1(b). The electric field gradient attracts dielectric objects using positive DEP, verified by measuring the pull-off force $F_{PO}$ of the CAT on a silicon substrate *vs.* the applied root-mean-square voltage $V$. We use the CAT to demonstrate three dimensional assembly by picking up a single targeted silica microsphere, imaging with the microsphere trapped at the tip of the CAT, and then depositing the microsphere at a desired location.

Dielectrophoresis is well suited for the manipulation of microscale objects. A dielectric sphere of radius $a$ in the presence of an applied electric field $\vec{E}$ will experience a dielectrophoretic force[9]

$$\vec{F}_{DEP} = 2\pi a^3 \varepsilon_M \, \text{Re}\left[\frac{\hat{\varepsilon}_P - \hat{\varepsilon}_M}{\hat{\varepsilon}_P + 2\hat{\varepsilon}_M}\right] \vec{\nabla} E^2, \tag{1}$$

where $\hat{\varepsilon}_P$ and $\hat{\varepsilon}_M$ are the complex permittivities of the sphere and medium, respectively, and $\varepsilon_P$ is the real permittivity of the object. The strength of $\vec{F}_{DEP}$ is only limited by the breakdown electric field $E_{BD}$ of the medium. For air nanogaps, $E_{BD} \sim 0.8$ MV/cm, dependent on electrode geometry.[12] For a silica microsphere with dielectric permittivity $\varepsilon_P = 3.8\varepsilon_0$ and radius $a = 500$ nm, at the tip of a CAT in air with $\varepsilon_M = \varepsilon_0$, we calculate $\vec{F}_{DEP} \lesssim 30$ to 35 nN from axisymmetric electrostatic simulation (Maxwell SV – Ansoft).[13]

Adhesion to surfaces, or the "sticky finger" problem, is a significant challenge for the manipulation of microscale objects. The Derjaguin, Muller, and Toporov (DMT) model of adhesion predicts an adhesion force $F_{DMT} \propto a$ between a spherical particle of radius $a$ and a plane.[14] For a silica microsphere with $a = 500$ nm in contact with a silica plane, $F_{DMT} \sim 200$ nN.[14] Because $F_{DEP}/F_{DMT} \propto a^2$, DEP can overcome adhesion for larger particles. Adhesion may also be mitigated by working in low humidity or on rough surfaces.[15]

We construct coaxial AFM tweezers as shown schematically in Fig. 1(c). We begin with a commercial conducting AFM probe (Arrow-NC - NanoAndMore USA) and make ohmic contact to the conducting probe by thermal evaporation of 15 nm of Ti and 100 nm of Al then annealing for 30 minutes at 325 °C in forming gas. An insulating low stress $SiN_x$ film is grown on the probe side via plasma enhanced chemical vapor deposition (Cirus 150 - Nexx), followed by thermal evaporation of a metallic shell composed of 30 nm of Cr and 30 nm of as shown in Fig. 1(c). The coaxial tip is created by etching with a Zeiss NVision 40 focused ion beam (FIB). Typical $SiN_x$ films had a thickness of 520 nm and a breakdown field $E_{BD}$ = 4.8±0.3 MV/cm. Coaxial probes are mounted on a holder with separate electrical contacts for the inner conductor and the ground shield in an Asylum MFP 3D AFM. The cantilever spring constant is measured using the equipartition technique.[1]

We show that DEP provides an attractive force towards the coaxial tip by measuring the pull-off force $F_{PO}$ from a substrate. Figure 2(a) shows the force $F$ on the tip *vs.* tip-sample distance $d$ with no applied voltage, and Fig. 2(b) shows $F$ *vs.* $d$ for $V = 20$ V at 5 kHz. The red and black lines show $F$ measured while descending and ascending as indicated. When the tip hits the sample, the force increases linearly until a set value at which point the tip is withdrawn. The largest attractive force is denoted the pull-off force $F_{PO}$. Figure 2(c) shows $F_{PO}$ *vs.* $V$; each 'x' represents an individual $F$ *vs.* $d$ measurement and each 'o' represents the average $F_{PO}$ at that voltage $V$. The red line is a fit to $F_{PO} = \beta + \alpha V^2$, where $\beta = 259 \pm 9$ nN represents adhesion and $\alpha = 2.10 \pm 0.04$ nN/V$^2$ is due to DEP. These results agree with axisymmetric simulations of similar coaxial probes which predict $\alpha =$ 2.3 to 2.8 nN/V$^2$.

We use silica beads on silicon wafers as a model system to demonstrate imaging and assembly. Polished silicon wafers (Silicon Quest International) were used as a substrate. Some wafers were roughened to reduce adhesion by etching through a layer of photoresist with a reactive ion etch to increase the rms roughness from 0.2 nm to 2.3 nm.[16] Cleaned silica beads (Polysciences, Inc.) of radius $a = 500$ nm were deposited on the substrate by pipetting on a 20 µL droplet of microspheres suspended in deionized water and allowing the solvent to evaporate in a nitrogen environment. The microspheres are manipulated with relative humidity below 20% to minimize capillary adhesion.

We demonstrate three dimensional assembly of silica beads using the CAT. Figures 3(a) to 3(c) show the manipulation of silica beads on polished silicon and Figs. 3(d) to 3(f) the manipulation of a close-packed monolayer of beads on rough silicon, both in air. To locate a bead, the sample is first imaged in tapping mode, as shown in Figs. 3(a) and 3(d). The triangular shape is due to the convolution of the triangular tip with the microspheres. To pick up a microsphere, the tip is pressed into it with $V = 40$ V at 5 kHz. By examining the force-distance curve, it is evident if a microsphere was successfully picked up. Once the microsphere has been picked up, the field is turned off, and the beads are imaged again as shown in Figs. 3(b) and 3(e). The microspheres now appear round, because they are imaged by a fellow microsphere. The microsphere is deposited by pressing the tip into the desired location and moving it along the surface. The bead is scraped off and remains at that location. If the drop off was successful, it will be evident from the AFM feedback immediately after the drop off. The substrate is again imaged as shown in Figs. 3(c) and 3(f) to verify that the move was successful.

The authors thank David Issadore, Jesse Berezovsky, Tom Hunt, Ania Bleszynski, and Noah Clay for valuable advice, Jim MacArthur for help with electronics, and Maarten Rutgers at Asylum Research for providing the tip holder. We acknowledge support by the Department of Defense through a National Defense Science &

FIG. 1. (Color Online) (a) Scanning electron micrograph of a coaxial AFM probe. The probe consists of a Cr/Au ground shield, SiNx insulator and n-doped Si inner conductor. (b) Quasistatic simulation of the electric field (Maxwell 11 – Ansoft) for the coaxial probe shown in (a), with the inner conductor held at 40 V and the outer shell grounded. (c) Schematic of the fabrication process showing deposition of a SiNx insulating layer followed by a Cr/Au conducting shell on the conducting AFM tip. The inner conductor is exposed by a focused ion beam (FIB) etch. Ohmic contact is made to the inner conductor by thermal evaporation of Ti/Al and subsequent annealing.

FIG. 2. (Color Online) Force $F$ on the tip *vs.* tip-sample distance $d$ with (a) no voltage applied and (b) $V = 20$ V$_{RMS}$ applied to the inner conductor at 5 kHz. The red line indicates the tip is moving toward the surface and the black line indicates that the tip is moving away from the surface. The largest attractive force during withdrawal is denoted the pull-off force $F_{PO}$. (c) $F_{PO}$ *vs.* $V$ where each 'x' is from a unique $F$-$d$ curve and each 'o' represents the average $F_{PO}$ at a given $V$. The red line is a fit of the averages to $F_{PO} = \beta + \alpha V^2$.

FIG. 3. (Color Online) Tapping mode images taken by the CAT showing the assembly of silica microspheres. The color scale corresponds to the change in oscillation

amplitude $\Delta H$. The scale bar is 2 μm. (a) Sparse microspheres on a smooth silicon substrate. The microsphere marked with an '*' is picked up by the CAT. (b) The same sample is imaged again with the microsphere on the tip of the CAT. (c) The microsphere is deposited in the location marked with an '*' and the sample is imaged again with the coaxial tip. (d) A close-packed layer of microspheres on a rough silicon substrate. The microsphere marked with an '*' is picked up with the CAT. (e) The same sample is imaged again with the microsphere on the tip. (f) The microsphere is deposited in the location marked with an '*' and the sample is imaged again with the coaxial tip. This microsphere was deposited on top of an array of other microspheres demonstrating 3D assembly.

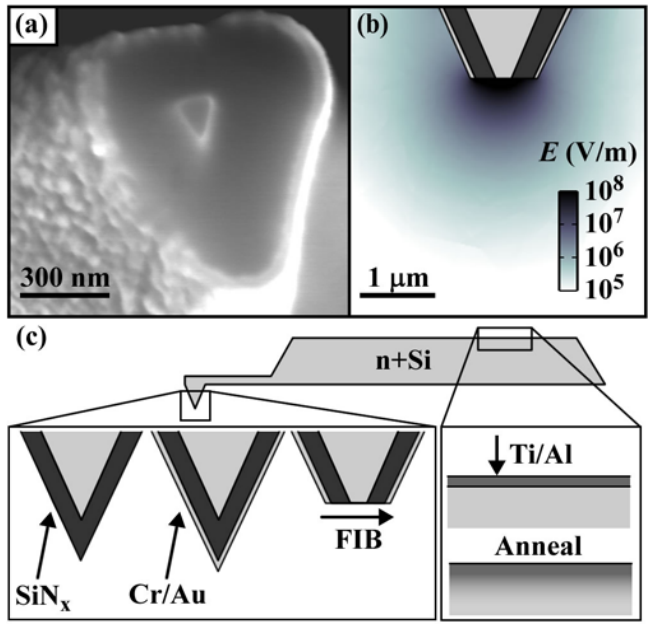

FIG.1.

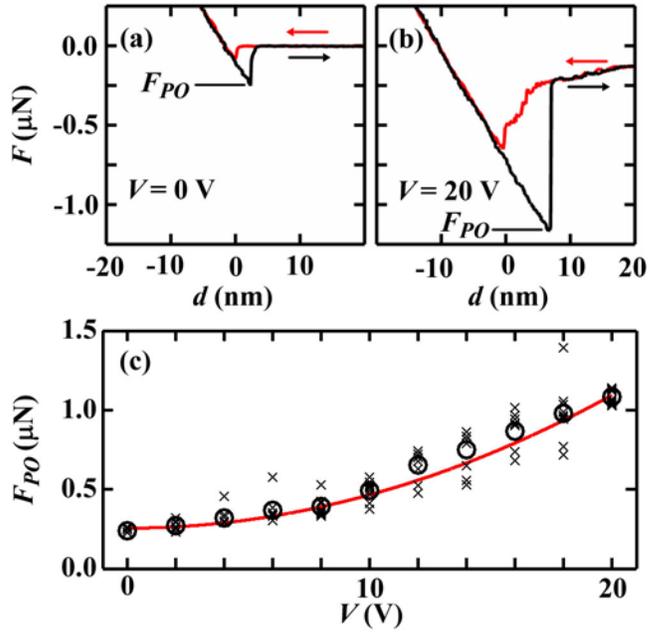

FIG. 2.

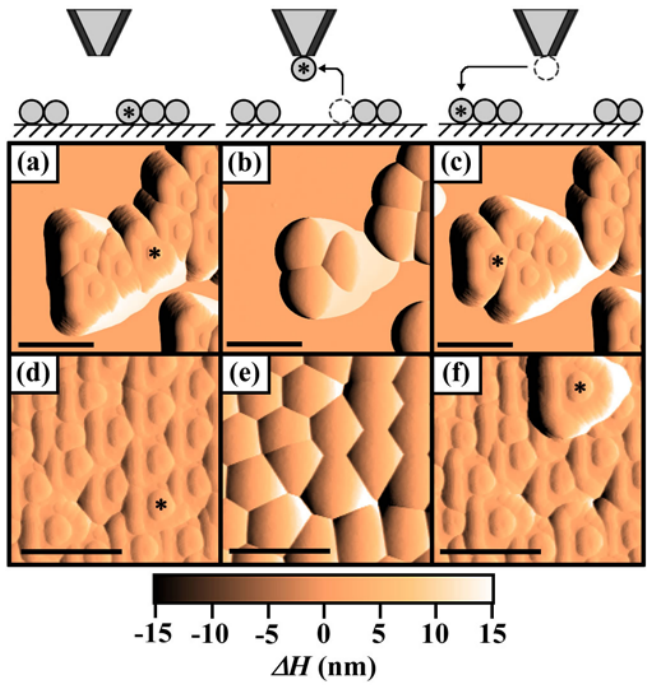

FIG.3